\newif\ifbio
\biotrue

\ifbio

\let\latexarabic\arabic
\let\latexdocument\document
\let\latexenddocument\enddocument

\RequirePackage[thmmarks]{ntheorem}
\makeatletter
\renewtheoremstyle{plain} 
  {\item[\hskip\labelsep \theorem@headerfont ##1\ \textup{##2}\theorem@separator]} 
  {\item[\hskip\labelsep \theorem@headerfont ##1\ \textup{##2}\ (##3)\theorem@separator]}
\makeatother

\documentclass[article]{biometrika}

\let\document\latexdocument
\let\enddocument\latexenddocument
\AtEndDocument{\printhistory}
\let\arabic\latexarabic
\def\rm{}

\usepackage{amsmath, amssymb, amsxtra, graphicx, float, caption, subcaption, color, paralist}

\usepackage{times}  
\usepackage{bm}
\usepackage{natbib}

\usepackage[plain,noend]{algorithm2e}

\makeatletter
\renewcommand{\algocf@captiontext}[2]{\quad #1\algocf@typo. \AlCapFnt{}#2} 
\def\@algocf@capt@plain{top}
\renewcommand{\algocf@makecaption}[2]{%
  \addtolength{\hsize}{\algomargin}%
  \sbox\@tempboxa{\algocf@captiontext{#1}{#2}}%
  \ifdim\wd\@tempboxa >\hsize
    \hskip .5\algomargin%
    \parbox[t]{\hsize}{\algocf@captiontext{#1}{#2}}
  \else%
    \global\@minipagefalse%
    \hbox to\hsize{\box\@tempboxa}
  \fi%
  \addtolength{\hsize}{-\algomargin}%
}
\makeatother

\newcommand{\p}[1]{\left(#1\right)}
\newcommand{\sqb}[1]{\left[#1\right]}
\newcommand{\cb}[1]{\left\{#1\right\}}
\newcommand{\EE}[2][]{E_{#1}\left(#2\right)}
\newcommand{\PP}[2][]{pr_{#1}\left(#2\right)}
\newcommand{\RR}{\mathbb{R}}
\newcommand{\htau}{\hat{\tau}}
\newcommand{\cond}{\mid}

\else 
\documentclass{article}
\usepackage[margin=1.5in]{geometry}
\usepackage{amsmath, amsthm, amssymb}
\usepackage{graphicx}
\usepackage{verbatim}
\usepackage{natbib}
\usepackage{caption}
\usepackage{subcaption}
\usepackage{fancyvrb}
\usepackage{enumerate}
\usepackage{relsize}
\usepackage{mathtools}
\usepackage{tikz}
\usepackage{float}
\usepackage{hyperref}
\hypersetup{colorlinks,citecolor=blue,urlcolor=blue, linkcolor=blue}
\usepackage{stefan_tex}

\theoremstyle{plain}

\newtheorem{proposition}{Proposition}
\newtheorem{theorem}[proposition]{Theorem}

\theoremstyle{definition}
\newtheorem{example}{Example}
\newtheorem{remark}{Remark}
\newtheorem{definition}{Definition}
\newtheorem{assumption}{Assumption}

\fi

\newcommand{\HH}{\operatorname{HH}}
\newcommand{\AMR}{\operatorname{AMR}}

\newcommand{\ADE}{\operatorname{ADE}}
\newcommand{\AIE}{\operatorname{AIE}}
\newcommand{\AOE}{\operatorname{AOE}}
\newcommand{\NUDGE}{\operatorname{INF}}
\newcommand{\IE}{\operatorname{IE}}

\newif\ifyuchen
\yuchenfalse

\begin{document}

\ifbio
\jname{Biometrika}
\jyear{}
\jvol{forthcoming}
\jnum{}




\author{Yuchen Hu}
\affil{Dept.~of Management Science and Engineering,
475 Via Ortega, Stanford, CA-94305, U.S.A. \email{yuchenhu@stanford.edu}}

\author{Shuangning Li}
\affil{Department of Statistics, 390 Jane Stanford Way, Stanford, CA-94305, U.S.A.
\email{lsn@stanford.edu}}

\author{\and Stefan Wager}
\affil{Stanford Graduate School of Business, 655 Knight Way, Stanford, CA-94305, U.S.A. 
\email{swager@stanford.edu}}

\else
\author{
        Yuchen Hu \and
        Shuangning Li \and
        Stefan Wager
}
\date{Stanford University}
\fi

\title{Average Direct and Indirect Causal Effects \\ under Interference}

\maketitle

\begin{abstract}
We propose a definition for the average indirect effect of a binary treatment in the potential outcomes model for causal inference under cross-unit interference. Our definition is analogous to the standard definition of the average direct effect, and can be expressed without needing to compare outcomes across multiple randomized experiments. We show that the proposed indirect effect satisfies a decomposition theorem whereby, in a Bernoulli trial, the sum of the average direct and indirect effects always corresponds to the effect of a policy intervention that infinitesimally increases treatment probabilities. We also consider a number of parametric models for interference, and find that our non-parametric indirect effect remains a natural estimand when re-expressed in the context of these models.
\end{abstract}

\ifbio
\fi

\section{Introduction}
\label{section:introduction}

The classical way of analyzing randomized trials, following \citet{neyman1923applications}
and \citet{rubin1974estimating}, is centered around the average treatment effect as defined using potential outcomes.
Given a sample of $i = 1, \, \ldots, \, n$ units used to study the effect of a binary treatment $W_i \in \cb{0, \, 1}$,
we posit potential outcomes $Y_i(0), \, Y_i(1) \in \RR$ corresponding to the outcome we would have measured
had we assigned the $i$-th unit to control or treatment respectively, i.e., we observe $Y_i = Y_i(W_i)$.
We then proceed by arguing that the sample average treatment effect
\begin{equation}
\label{eq:ATE}
\tau_{ATE} = \frac{1}{n} \sum_{i=1}^n \cb{Y_i(1) - Y_i(0)}
\end{equation}
admits a simple unbiased estimator under random assignment of treatment.

One limitation of this classical approach is that it rules out interference, and instead introduces an assumption that
the observed outcome for any given unit does not depend on the treatments assigned to other units, i.e.,
$Y_i$ is not affected by $W_j$ for any $j \neq i$ \citep{halloran1995causal}. However, in a wide variety of applied settings,
such interference effects not only exist, but are often of considerable scientific interest
\citep{bakshy2012role,bond201261,cai2015social,miguel2004worms,rogers2018reducing,sacerdote2001peer}.
For example, in an education setting, it may be of interest to understand how a didactic innovation affects
not only certain targeted students, but also their peers.
This has led to a recent surge of interest in methods for studying randomized trials under interference
\citep{aronow2017estimating,eckles2017design,hudgens2008toward,leung2020treatment,
li2020random,manski2013identification,savje2017average,tchetgen2012causal}.

A major difficulty in working under interference is that we no longer have a single obvious average
effect parameter to target as in \eqref{eq:ATE}. In the general setting, each unit now has $2^n$ potential outcomes
corresponding to every possible treatment combination assigned to the $n$ units, and these can be used to formulate
effectively innumerable possible treatment effects that can arise from different assignment patterns.
As discussed further in Section \ref{sec:relwork} below, the existing literature has mostly side-stepped this issue by
framing the estimand in terms of specific policy interventions. However, this paradigm does not provide researchers
with simple, non-parametric and agnostic average causal estimands that can be studied without spelling out a specific policy intervention of interest.

In this paper, we study a pair of averaging causal estimands, the average direct and indirect effects,
that are valid under interference yet, unlike existing targets, can be defined and estimated using a single experiment
and do not need to be defined in terms of hypothetical policy interventions. Qualitatively, the average direct effect measures the extent to which, in a given experiment and on average, the outcome $Y_i$ of a unit is affected by its own
treatment $W_i$; meanwhile, the average indirect effect measures the responsiveness of $Y_i$ to treatments $W_j$ given to other units $j \neq i$.

The average direct effect we consider is standard and has recently been discussed by a number of authors including \citet{vanderweele2011effect} and \citet*{savje2017average}. Our definition
of the average indirect effect is to the best of our knowledge new, and is the main contribution of this paper.
We follow this definition with a number of results to validate it. In particular, we prove a universal decomposition theorem
whereby, in a Bernoulli trial, the sum of the average direct and indirect effects can always be interpreted as the total effect of an intuitive
policy intervention. We also interpret these estimands in the context of a number of parametric models for interference
considered by practitioners.


\section{Treatment Effects under Interference}
\label{sec:main}


We study different experimental designs using the potential outcomes model.
The main difference between a setting with interference and the standard Neyman-Rubin model is that 
potential outcomes for the $i$-th unit may also depend on the intervention given to the $j$-th unit with $j \neq i$
\citep[e.g.,][]{aronow2017estimating,hudgens2008toward}.
For convenience, we use short-hand $Y_i(w_j = x; \, W_{-j})$ to denote the potential outcome we
would observe for the $i$-th unit if we were to assign the $j$-th unit to treatment status $x \in \cb{0, \, 1}$, and
otherwise maintained all but the $j$-th unit at their realized treatments
$\smash{W_{-j} \in \cb{0, \, 1}^{n-1}}$.
Expectations $E$ are over the treatment assignment
only; potential outcomes are held fixed.

\begin{assumption}
\label{assu:PO}
For units $i = 1, \, \ldots, \, n$, there are potential outcomes \smash{$Y_i(w) \in \RR$},
\smash{$w \in \cb{0, \, 1}^n$} such that, given a treatment vector
\smash{$W \in \cb{0, \, 1}^n$}, we observe outcomes \smash{$Y_i = Y_i(W)$}.
\end{assumption}

\begin{definition}
\label{defi:ADE}
Under Assumption \ref{assu:PO}, the average direct effect of a binary treatment
is 
\begin{equation}
\label{eqn:ADE}
\tau_{\ADE} = \frac{1}{n} \sum_{i=1}^n E\cb{Y_i(w_i = 1; W_{-i}) - Y_i(w_i = 0; W_{-i})}. 
\end{equation}
\end{definition}

\begin{definition}
\label{defi:AIE}
Under Assumption \ref{assu:PO}, the average indirect effect of a binary treatment
is
\begin{equation}
\label{eqn:AIE}
\tau_{\AIE} = \frac{1}{n}\sum_{i = 1}^n \sum_{j \neq i}  E \cb{Y_j(w_i=1;W_{-i})-Y_j(w_i=0;W_{-i}) }.
\end{equation}
\end{definition}

The definition of the direct effect $\tau_{\ADE}$ is standard. It follows from averaging $Y_i(w_i = 1; W_{-i}) - Y_i(w_i = 0; W_{-i})$, referred to as the direct causal effect by \citet{halloran1995causal}.
\citet{savje2017average} provide a recent in-depth discussion on this estimand.
This estimand measures the average effect of an intervention $W_i$
on the unit being intervened on---while marginalizing over the rest of the treatment assignments.
In a study without interference, $\tau_{\ADE}$ matches the usual average treatment effect \eqref{eq:ATE}.

Meanwhile, our definition of the indirect effect is an immediate formal generalization of $\tau_{\ADE}$ to cross-unit
treatment effects. It measures the average effect of an intervention $W_i$ on all units except the one being
intervened on, again marginalizing over the rest of the process. More precisely, the term $E\cb{Y_j(w_i=1;W_{-i})-Y_j(w_i=0;W_{-i}) }$ is the effect of changing unit $i$'s treatment on the outcome of unit $j$. Thus the sum, $\sum_{j \neq i}  E\cb{Y_j(w_i=1;W_{-i})-Y_j(w_i=0;W_{-i}) }$, would correspond to the aggregate effect of unit $i$'s treatment on all other units. Then the defined average indirect effect $\tau_{\AIE}$ corresponds to the average of the effects of units' treatments on other units.

The definition of $\tau_{\AIE}$ formally mirrors that of $\tau_{\ADE}$, and in the no-interference case we clearly have
$\tau_{\AIE} = 0$.
As a first step towards validating the definition of $\tau_{\AIE}$ under non-trivial interference,
we consider the average overall effect induced by adding
$\tau_{\ADE}$ and $\tau_{\AIE}$, which aggregates the marginalized effect of all treatments on all outcomes.
We then prove that, in a Bernoulli design, this matches the policy effect of infinitesimally increasing
each unit's treatment probability.
We use the term Bernoulli design to refer to an experiment where there is
a deterministic vector $\pi \in (0, \, 1)^n$ such that the treatments $W_i$
are generated as
$W_i \, \sim \, \operatorname{Bernoulli}(\pi_i)$ for all $i = 1, \, \ldots, \, n$,
independently of each other and of the potential outcomes $\cb{Y_i(w)}$.
For a Bernoulli design with treatment probabilities $\pi \in [0, \, 1]^n$, we write
$\EE[\pi]{\cdot}$ for expectations over the random treatment assignment, and write $\tau_{\ADE}(\pi)$, $\tau_{\AIE}(\pi)$ and
$\tau_{\AOE}(\pi)$ for the corresponding direct, indirect and overall effects.

\begin{definition}
\label{defi:AOE}
Under Assumption \ref{assu:PO}, the average overall effect of a binary treatment is
\begin{align}
\label{eqn:AOE}
\tau_{\AOE} = \tau_{\ADE} + \tau_{\AIE} =\frac{1}{n}\sum_{i = 1}^n \sum_{j=1}^n  E\cb{Y_j(w_i=1;W_{-i})-Y_j(w_i=0;W_{-i}) }.
\end{align}
\end{definition}


\begin{definition}
\label{defi:NUDGE}
Under Assumption \ref{assu:PO} and in a Bernoulli design, the infinitesimal policy effect is 
\begin{equation}
\label{eqn:NUDGE}
\tau_{\NUDGE}(\pi) = \mathbf{1} \cdot \nabla_{\pi} \EE[\pi]{\frac{1}{n}\sum_{i = 1}^n Y_i} = \sum_{k=1}^n \frac{\partial}{\partial \pi_k} \EE[\pi]{\frac{1}{n}\sum_{i = 1}^n Y_i}. 
\end{equation}
\end{definition}

\begin{theorem}
\label{theo:overall}
Under Assumption \ref{assu:PO} and in a Bernoulli design,
\smash{$\tau_{\AOE}(\pi) = \tau_{\NUDGE}(\pi)$}. 
\end{theorem}

By connecting our abstract notions of direct, indirect and overall effects to the effect of a concrete policy intervention,
Theorem \ref{theo:overall} provides an alternative lens on our definition of the indirect effect. Suppose, for example,
that a researcher knew they wanted to study nudge interventions, the total effect of which is $\tau_{\NUDGE}(\pi)$,
and was also committed to the standard definition of the average direct effect given in Definition \ref{defi:ADE}.
Then, it would be natural to define an indirect effect as $\tau_{\NUDGE}(\pi) - \tau_{\ADE}(\pi)$, i.e., to characterize
as indirect effect any effect of the nudge intervention that is not captured by the direct effect; this is, for example,
the approach implicitly taken in \citet{heckman1998general}. From this perspective, Theorem \ref{theo:overall}
could be seen as showing that these two possible definitions of the indirect effect in fact match, i.e., that 
$\tau_{\AIE}(\pi) = \tau_{\NUDGE}(\pi) - \tau_{\ADE}(\pi)$. We emphasize that Theorem \ref{theo:overall} is a direct
consequence of Bernoulli randomization, and holds conditionally on any realization of the potential outcomes $\cb{Y_i(w)}$.

We refer to $\tau_{\NUDGE}(\pi)$ as a policy effect because,
in the ideal situation when one has access to observed outcomes $Y_i$ for different randomization probabilities $\pi$,
$\tau_{\NUDGE}(\pi)$ is a quantity that could be measured by
averaging observed outcomes $Y_i$ for different $\pi$.
If treatment assignment probabilities are constant, i.e., there is a $\pi_0 \in (0, \, 1)$ such that $\pi_i = \pi_0$ for all $i = 1, \, \ldots, \, n$, then $\tau_{\NUDGE}(\pi)$ takes on a particularly simple form,
\smash{$\tau_{\NUDGE}(\pi) =  {d}/{d\pi_0} \, \EE[\pi_0]{n^{-1} \sum_{i=1}^n Y_i}$}. 
Infinitesimal policy effects as defined above
are prevalent in social sciences
due to their ease of interpretation and desirable analytic properties; see, e.g., \citet{chetty2009sufficient}, \citet{carneiro2010evaluating},
and references therein. \citet{wager2021experimenting} discuss welfare implications for a social planner who
uses analogous infinitesimal policy effects to optimize a system via gradient-based methods.

\begin{remark}
Accurate estimation of $\tau_{\ADE}$ and $\tau_{\AIE}$ is an interesting
question beyond the scope of this paper. In Appendix
\ref{sec:unbiased}, we show that both $\tau_{\ADE}$ and $\tau_{\AIE}$ 
always admit unbiased estimators in Bernoulli experiments.
However, the precision of these unbiased estimators
will depend on the interference pattern. \citet{li2020random}
study estimators for $\tau_{\ADE}$ and $\tau_{\AIE}$ in the context of a
random graph model for interference, including improvements in settings where simple
unbiased estimators are unstable.
\end{remark}

\section{Alternative Definitions and Related Work}
\label{sec:relwork}

There are also a number of other average causal effect estimands that have recently been discussed in the literature.
In the case of direct effects, the main alternative to Definition \ref{defi:ADE} is the following proposal from
\citet{hudgens2008toward} that relies on conditional expectations,
\begin{equation*}
\label{eqn:HH_ADE}
\tau_{\HH, \operatorname{DE}} =  \frac{1}{n} \sum_{i=1}^n \cb{\EE{Y_i \cond W_i = 1} - \EE{Y_i \cond W_i = 0}}. 
\end{equation*}
In a Bernoulli design, $\tau_{\HH, \operatorname{DE}} = \tau_{\ADE}$. However, in other designs, e.g., completely
randomized designs or stratified designs, these two estimands do not match. 
As discussed in \citet{vanderweele2011effect} and \citet{savje2017average}, a major drawback of the definition $\tau_{\HH, \operatorname{DE}}$
is that it conflates the effect of setting $w_i = x$ on the $i$-th unit's outcome, and the effect of setting $w_i = x$ on
the distribution of $W_{-i}$. In particular, in completely randomized experiments, it's possible to have
$\tau_{\HH, \operatorname{DE}} \neq 0$ even when $Y_i(w_i = 1, \, w_{-i}) = Y_i(w_i = 0, \, w_{-i})$ for
all units and all possible treatment assignments. In contrast, $\tau_{\ADE}$ as defined in Definition \ref{defi:ADE}
has a robust causal interpretation as a direct effect.

Meanwhile, as discussed in the introduction, most available notions of indirect effects rely on explicit
comparisons between two overall treatment assignment strategies.
For example, \citet{hudgens2008toward} and \citet{vanderweele2011effect} propose a number of
indirect effect estimands that, in the case of comparing two Bernoulli trials with randomization probabilities $\pi$ and $\pi'$, reduce to
\begin{equation*}
\label{eqn:HH}
\tau_{\IE}(\pi, \, \pi') = \frac{1}{n} \sum_{i = 1}^{n} \sqb{E_\pi\cb{Y_{i}(w_i = 0 ; W_{-i})} - E_\pi\cb{Y_{i}(w_i = 0; W_{-i})}}.
\end{equation*}
In the case of non-Bernoulli trials, there are a number of subtleties analogous to the ones noted above; see
\citet{vanderweele2011effect} for an in-depth discussion. \smash{$\tau_{\IE}(\pi, \, \pi')$}
is an interesting quantity to consider if we can run many independent experiments that test different overall treatment level but,
unlike $\tau_{\AIE}$, does not enable a researcher to describe indirect effects in a single randomized study.

Another popular approach to capturing indirect effects is via the exposure mapping approach developed in \citet{aronow2017estimating}.
The main idea is to assume existence of functions $h_i : \cb{0, \, 1}^n \rightarrow \cb{1, \, \ldots, \, K}$ such that potential
outcomes $Y_i(w)$ only depend on $w$ via the compressed representation $h_i(w)$, i.e., $Y_i(w) = Y_i(w')$ whenever
$h_i(w) = h_i(w')$; see also \citet{karwa2018systematic}, \citet{leung2020treatment}
and \citet{savje2021causal} for further discussions and extensions.
One can then consider estimators of
averages of potential outcome types and define treatment effects in terms of their contrasts,
\begin{equation*}
\label{eq:AS}
\mu(k) = \frac{1}{n} \sum_{i = 1}^n E\cb{Y_i \cond h_i(W_i) = k}, \ \ \ \ \ \tau(k, \, k') = \mu(k') - \mu(k), \ \ \ \ \ 1 \leq k \neq k' \leq K.
\end{equation*}
Definitions of this type are again conceptually attractive and sometimes enable us to very clearly express
the answer to a natural policy question; see, e.g., \citet{basse2019randomization}. However, they again require the
analyst to consider specific policy interventions to be able to even talk about indirect effects, and can also be unwieldy
to use as the number of possible exposure types $K$ gets large.

Closest to the definition of $\tau_{\AIE}$ is the average marginalized response of \citet{aronow2020design}.
They consider a setting where treatments are assigned to points in a geographic space, and seek to estimate
the average effect of treatment at an intervention point outcomes at points that are a distance $d$ away,
\begin{equation*}
\label{eqn:AMR}
\tau_{\AMR}(d;\pi) = \frac{1}{n} \sum_{i = 1}^{n} \sum_{j \in \mathcal{S}_i(d)} \frac{E_{\pi}\cb{Y_{j}(w_i = 1 ; W_{-i}) - Y_{j}(w_i = 0; W_{-i})}}{|\mathcal{S}_i(d)|}, 
\ \ \mathcal{S}_i(d) = \{j : \Delta(i, \, j) = d\},
\end{equation*}
where $\Delta(i, \, j)$ measures the distance between points $i$ and $j$.
This circle average bears resemblance to Definition \ref{defi:AIE} in the sense that both of them are marginalized
over variation in $W_{-i}$ holding the treatment $W_i$ fixed. However, one key difference is the normalization factor
$|\mathcal{S}_i(d)|^{-1}$ used in $\tau_{\AMR}$. Adding similar normalization to $\tau_{\AIE}$ would invalidate Theorem \ref{theo:overall}.

\begin{remark}
Our definition of $\tau_{\AIE}(\pi)$ is normalized by $n$, not
by the total number of summands $n(n-1)$, and one can ask whether this scaling is always the most natural one.
We argue below that, in a number of popular models,
$\tau_{\AIE}(\pi)$ coincides with interesting and interpretable quantities and converges as the number of units $n$ goes to infinity. 
In other models, however, our $1/n$ scaling may not be the best choice. For example, 
if we have a data-generating distribution with $Y_i(w) = (\sum_{j = 1}^n w_j - n \pi) / \sqrt{n \pi (1 - \pi)}$,
the observed outcomes $Y_i$ will have a standard normal marginal distribution,
but $\tau_{\AIE} = \sqrt{n}$ diverges. One should also note, however, that in this example $1/n \sum_{i = 1}^n Y_i$ does not concentrate.
\end{remark}

\section{Models for Interference}
\label{sec:models}

Our discussion so far has focused on an abstract specification where direct and indirect effects are defined
via various marginalized contrasts between potential outcomes. Much of the existing applied work on
causal inference under interference, however, has focused on simpler parametric specifications that, e.g.,
connect outcomes to treatments via a linear model. The purpose of this section is to examine our abstract,
non-parametric definition of the indirect effect given in Definition \ref{defi:AIE}, and to confirm that it still
corresponds to an estimand one would want to interpret as an indirect effect once we restrict our attention
to simpler parametric models. Below, we do so in 3 examples. \citet{munro2021treatment} provide an extended
study of $\tau_{\ADE}$ and $\tau_{\AIE}$ in a marketplace model where interference arises via equilibrium price formation.
The claimed expressions for $\tau_{\ADE}$ and $\tau_{\AIE}$ are derived in Appendix \ref{section:proofs}. 

\begin{example}
\label{exam:1}
In studying the spillover effects of insurance training sessions on insurance purchase, \citet{cai2015social}
work in terms of a network model: There is an edge matrix $E_{ij} \in \cb{0, \,1}$, such that $W_j$ can only
affect $Y_i$ if the corresponding units are connected by an edge, i.e., if $E_{ij} = 1$. They then consider a
linear-in-means model parametrized in terms of this network. For our purpose,
we focus on a simple variant of the model of \citet{cai2015social} considered in \citet{leung2020treatment},
where only the effects of ego's treatment and the proportion of treated neighbors are considered as covariates.
This results in a linear model induced by the structural equation
\begin{equation}
\label{exam1}
Y_i = \beta_1+\beta_2 W_i +\beta_3 \frac{\sum_{j\ne i} E_{ij}W_j}{\sum_{j\ne i} E_{ij}} + \varepsilon_i, \ \ \EE{\varepsilon_i \cond W} = 0.
\end{equation}
In words, the probability of insurance purchase is modeled as a linear function of whether the farmer attends the insurance training sessions, and the proportion of friends who attend the session. The relation \eqref{exam1} should be taken as a structural model, meaning that we can generate potential outcomes $Y_i(w)$ by plugging candidate assignment vectors $w$ into \eqref{exam1}, i.e., \smash{$Y_i(w) = \beta_1 + \beta_2 w_i + \beta_3 \sum_{j \neq i} E_{ij} w_j / \sum_{j \neq i} E_{ij} + \varepsilon_i$}, for all $w \in \cb{0, \, 1}^n$.
Under this model, it can be shown that under Assumption \ref{assu:PO},
$\tau_{\ADE} = \beta_2$, and $\tau_{\AIE}=\beta_3$,
i.e., the estimands from Definitions \ref{defi:ADE} and \ref{defi:AIE}
map exactly to the parameters in model \eqref{exam1} regardless of the experimental design. 
\end{example}

\begin{example}
\label{exam:2}
The model in Example \ref{exam:1} assumes the the $i$-th unit responds in the same way to treatment assigned to any of
its neighbors. However, this restriction may be implausible in many areas; for example, in social networks, there is evidence
that some ties are stronger than others, and that peer effects are larger along strong ties \citep{bakshy2012role}.
A natural generalization of Example \ref{exam:1} that allows for variable strength ties is to consider a saturated structural linear model
\begin{equation}
\label{exam2}
Y_i = \alpha_i + \beta_iW_i + \sum_{j\ne i}\nu_{ij}W_j  + \varepsilon_i, \ \ \EE{\varepsilon_i \cond W} = 0,
\end{equation}
which allows for both unit-specific direct and indirect effects. Here the individual parameters in this model are
not identifiable; however, under \eqref{exam2},
$\tau_{\ADE}=\frac{1}{n}\sum_{i=1}^n \beta_i$, and 
$\tau_{\AIE}=\frac{1}{n}\sum_{i=1}^n\sum_{j\ne i} \nu_{ij}$,
i.e., our estimands can be understood as averages of the unit-level parameters, again regardless of the design.
\end{example}

\begin{example}
\label{exam:3}
In studying the effect of persuasion campaigns or other types of messaging, one may assume
that people respond most strongly if they get a communication directly addressed to them, but can also
respond if a member of their neighborhood or their household gets a communication. This assumption can
be formalized in terms of the following model: Each unit has 4 potential outcomes defined as
\begin{equation}
\label{exam3}
Y_i = \begin{cases}
Y_i(\text{treated \& exposed}) & \text{if $W_i = 1$ and $i$ has a treated neighbor}, \\
Y_i(\text{treated}) & \text{if $W_i = 1$ but $i$ has no treated neighbors}, \\
Y_i(\text{exposed}) & \text{if $W_i = 0$ but $i$ has a treated neighbor}, \\
Y_i(\text{none}) & \text{if $W_i = 0$ and $i$ has no treated neighbors}.
\end{cases}
\end{equation}
Models of this type are considered by \citet{sinclair2012detecting} for studying voter mobilization, and by
 \cite{basse2018analyzing} and \citet{basse2019randomization} for studying anti-absenteeism interventions. 
Natural treatment effect parameters to consider following \eqref{eq:AS} include the average self-treatment
and spillover effects
\begin{alignat*}{2}
\label{eq:self_spill}
\tau_{\text{SELF},1} &= \frac{1}{n} \sum_{i = 1}^n \cb{Y_i(\text{treated \& exposed}) - Y_i(\text{exposed})}, \quad
\tau_{\text{SELF},0} &&= \frac{1}{n} \sum_{i = 1}^n \cb{Y_i(\text{treated}) - Y_i(\text{none})},\nonumber\\
\tau_{\text{SPILL},1} &= \frac{1}{n} \sum_{i = 1}^n \cb{Y_i(\text{treated \& exposed}) - Y_i(\text{treated})}, \quad
\tau_{\text{SPILL},0} &&= \frac{1}{n} \sum_{i = 1}^n \cb{Y_i(\text{exposed}) - Y_i(\text{none})}.
\end{alignat*}
Unlike the previous two examples, the connection between $\tau_{\ADE}$ and $\tau_{\AIE}$ to $\tau_{\text{SELF}}$ and $\tau_{\text{SPILL}}$ differs substantially across experimental designs, especially when the design introduces correlation between the units. For the purpose of illustration, we study it in a multi-stage completely randomized design considered in previous works \citep{sinclair2012detecting, basse2018analyzing, basse2019randomization}. In particular, we focus on the case where there are in total $n/m$ clusters of size $m$. On the first stage, $\rho\cdot n/m$ clusters are assigned to treatment, and $(1-\rho)\cdot n/m$ clusters are assigned to control; on the second stage, a single unit in each treated cluster is randomly chosen to be treated, and all the other units are assigned to control. We can then calculate the marginal distribution of $W_{-i}$ and obtain
\begin{equation}
\label{exam3_targets2}
\begin{split}
&\tau_{\ADE} = \p{\rho-\frac{\rho}{m}}\tau_{\text{SELF},1}+\p{1-\rho+\frac{\rho}{m}}\tau_{\text{SELF},0},\\
&\tau_{\AIE} = (m-1)\cb{\frac{\rho}{m}\tau_{\text{SPILL},1}+\p{1-\rho+\frac{\rho}{m}}\tau_{\text{SPILL},0} }.
\end{split}
\end{equation}
Therefore, our estimands can be regarded as weighted averages of the treatment effect parameters. Moreover, when the cluster size $m$ is large, $\tau_{\ADE}$ is approximately the average of $\tau_{\text{SELF},1}$ and $\tau_{\text{SELF},0}$, weighted by the assignment probability during the first stage, while  $\tau_{\ADE}$ is approximately $\tau_{\text{SPILL},0}$ times the probability of being assigned to the control group during the first stage, and the factor $m-1$ simply
accounts for the fact that, any treatment will spread spillover effects to $m-1$ neighbors.
\end{example}

\section{Discussion}
\label{sec:discussion}


There are many treatment effect estimation problems in which interference is present and needs to be accounted for, and indirect effects are of considerable scientific interest.
We have proposed an estimand, $\tau_{\AIE}$, that quantifies the typical effect of treating one unit on the outcomes of all other units,
and provides a natural counterpart to the average direct effect. Consider,
for example, an experiment seeking to measure the effect of a housing voucher on homeownership. Here,
$\tau_{\ADE}$ measures the marginal benefit of receiving such a voucher oneself, whereas $\tau_{\AIE}$
analogously measures the extent to which one person receiving a voucher crowds out other buyers.
Theorem \ref{theo:overall} validates this perspective, by showing that the sum of $\tau_{\ADE}$ and $\tau_{\AIE}$
matches the expected effect of giving out more housing vouchers on total homeownership.

Our definition of $\tau_{\AIE}$ also has the potential to help synthesize non-parametric and model-based approaches
to interference by providing a shared estimand that can be studied from both perspectives:
As discussed in Section \ref{sec:models}, while $\tau_{\AIE}$ is defined
in terms of a generic potential outcomes model, it is also a natural estimand in a number of different
structural models. \citet{munro2021treatment}
pursue this agenda further and show that, in a marketplace governed by a general equilibrium
model where prices mediate interference, $\tau_{\AIE}$ can be expressed in terms of familiar economic quantities such
as price elasticities.

\begin{figure}
\includegraphics[width=\textwidth]{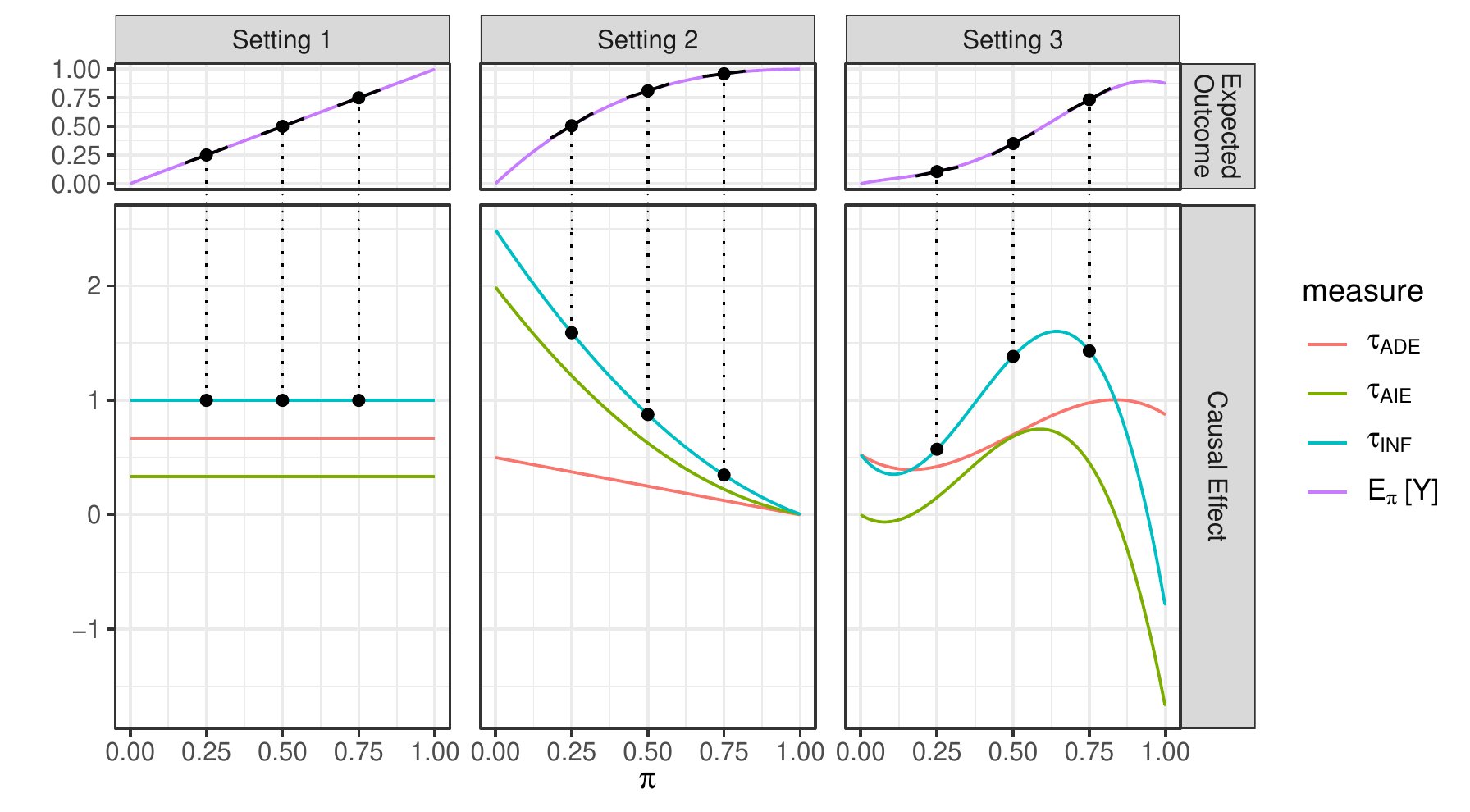}
\caption[LoF entry]{
Illustration of $\tau_{\ADE}$, $\tau_{\AIE}$, $\tau_{\NUDGE}$ and $\EE[\pi]{Y_i}$: The purple curve corresponds to the expected potential outcome $\EE[\pi]{Y_i}$. The slope of the line segments on the purple curve, i.e., the derivative of $\EE[\pi]{Y_i}$, is the same as the value of the blue curve ($\tau_{\NUDGE}$). Theorem \ref{theo:overall} establishes that $\tau_{\NUDGE} = \tau_{\ADE} + \tau_{\AIE}$. 
In the plots, the blue curve ($\tau_{\NUDGE}$) corresponds to the sum of the red curve and the green curve.  
We consider the three settings listed in \eqref{eq:settings} where, in all cases, we
assume constant treatment assignment probabilities $\pi_i = \pi_0$ and take the
number of neighbors to be $\sum_{i \neq j} E_{ij} = 100$. \vspace{-\baselineskip}}
\label{fig:curves}
\end{figure}

One challenge is that our estimands will in general depend on the design.
In Figure \ref{fig:curves}, we illustrate this phenomenon in a Bernoulli experiment by plotting $\tau_{\ADE}(\pi)$ and $\tau_{\AIE}(\pi)$
as a function of $\pi$ in the following three structural models with constant treatment probabilities $\pi$,
\begin{equation}
\label{eq:settings}
\begin{split}
&\text{\it 1:} \ \ Y_i = \frac{\sum_{i \neq j} E_{ij}W_j}{300} + \frac{2 W_i}{3} + \varepsilon_i, \ \ \ \ \ \ 
\text{\it 2:} \ \ Y_i = 1 - \p{1 - \frac{\sum_{i \neq j} E_{ij}W_j}{\sum_{i \neq j} E_{ij} }}^2 \p{1 - \frac{W_i}{2}} + \varepsilon_i, \\
&\text{\it 3:} \ \  Y_i = W_i\cb{e_i - 3\p{e_i - \frac{1}{2}}^3} + \varepsilon_i, \ \ \ e_i = \frac{\sum_{i \neq j} E_{ij}W_j}{\sum_{i \neq j} E_{ij} },
\end{split}
\end{equation}
where in each case $\EE{\varepsilon_i \cond W} = 0$. Here, qualitatively, Setting 1 resembles the one considered
by \citet{cai2015social} and \citet{leung2020treatment} as discussed in Example 1 from Section \ref{sec:models},
Setting 2 exhibits a type of herd immunity where units are more sensitive to treatment when most of their neighbors
are untreated, while Setting 3 has complicated non-linear interference effects. We then see that, in Setting 1,
$\tau_{\ADE}(\pi)$ and $\tau_{\AIE}(\pi)$ do not vary with $\pi$, but in Settings 2 and 3 they do---and may even
change signs.

\ifbio
\else
\captionsetup{singlelinecheck=off}
\fi

This potential dependence of $\tau_{\ADE}(\pi)$ and $\tau_{\AIE}(\pi)$ on $\pi$ is something that any
practitioner using these estimands needs to be aware of. However, we believe such dependence to be
largely unavoidable when seeking to define non-parametric estimands under the generality considered
here. For example, when estimating indirect effects of immunization in a population
where roughly 30\% of units have been immunized, definitions of the type developed here could be
used to support non-parametric analysis of indirect effects. Now, one should recognize that any such
effects would be local to the current overall immunization rate at 30\%, and would likely differ from indirect
effects we would measure at a 50\% overall immunization rate. However, it seems unlikely that one could use data from
a population with 30\% immunization rate to non-parametrically estimate average outcomes we might
observe at a 50\%; rather, to do so, one would need to either posit a model for how infections spread
or collect different data.


\ifbio
\bibliographystyle{biometrika}
\else
\bibliographystyle{plainnat}
\fi
\bibliography{references}

\newpage

\begin{appendix}

\section{Unbiased Estimation}
\label{sec:unbiased}

The main focus of this paper has been on establishing definitions of average treatment effect metrics under interference, and
on verifying their interpretability under various modeling assumptions. But in order for these definitions to be useful in
applied statistical work, we of course also need these metrics to be readily identifiable and estimable under flexible
conditions. A comprehensive discussion of treatment effect estimation under interference---including distributional
results and efficiency theory---is beyond the scope of this paper. However, as one result in this direction, we note
here that unbiased estimates of the average direct and indirect effects are always available in Bernoulli-randomized
experiments via the Horvitz-Thompson construction.

As above, the case of the average direct effect is already well understood in the literature. The Horvitz-Thompson
estimator for $\tau_{\ADE}(\pi)$ is \citep{savje2017average}
\begin{equation}
\label{eq:HT-ADE}
\htau_{\ADE}(\pi) = \frac{1}{n} \sum_{i = 1}^n \cb{\frac{W_i Y_i}{\pi_i} - \frac{(1 - W_i) Y_i}{1 - \pi_i}}.
\end{equation}
Furthermore, as is implicitly established in the technical appendix of \citet{savje2017average}, this estimator is
unbiased in Bernoulli-randomized experiments, i.e., \smash{$E\cb{\htau_{\ADE}(\pi)} = \tau_{\ADE}(\pi)$}.

Next, in discussing estimators for $\tau_{\AIE}(\pi)$, we will work under a
network interference model whereby the analyst has access to an interference graph $E_{ij} \in \cb{0, \, 1}$
and knows that the $j$-th unit's potential outcomes are unaffected by the treatment given to the $i$-th unit
whenever $E_{ij} = 0$, i.e.,
\begin{equation}
\label{eq:network}
Y_j(w_i = 0; \, w_{-i}) = Y_j(w_i = 1; \, w_{-i}) \text{   whenever   } E_{ij} = 0,
\end{equation}
for all $i,j = 1, \, \ldots, \, n$ and $w_{-i} \in {0, \, 1}^{n-1}$. This type of assumption is not required in principle; and
in particular, the condition \eqref{eq:network} is vacuous if $E_{ij} = 1$ for all pairs $(i, \, j)$, i.e., if we assume that
any treatment can affect any outcome. However, assumptions of this type are ubiquitous in practice, see, e.g.,
Examples \ref{exam:1} and \ref{exam:3} considered in Section \ref{sec:models}, and when we have access to a sparse interference graph
they can considerably improve the precision with which we can estimate indirect effects.

Given this setup, the Horvitz-Thompson estimator for $\tau_{\AIE}(\pi)$ is
\begin{equation}
\label{eq:HT-AIE}
\htau_{\AIE}(\pi) = \frac{1}{n} \sum_{i = 1}^n \sum_{\cb{j \neq i \, : \, E_{ij} = 1}} \cb{\frac{W_i Y_j}{\pi_i} - \frac{(1 - W_i) Y_j}{1 - \pi_i}}.
\end{equation}
Formally, this estimator looks like \smash{$\htau_{\ADE}(\pi)$}; except now we are measuring associations between the
$i$-th unit's treatment and the $j$-th unit's outcome, for $i \neq j$. And, as in the case of the direct effect, this estimator
in unbiased in Bernoulli-randomized experiments.

\begin{theorem}
\label{theo:unbiased}
Under Assumption \ref{assu:PO} and in a Bernoulli trial, the Horvitz-Thompson estimator \eqref{eq:HT-AIE} is unbiased
for the average indirect effect, \smash{$E\cb{\htau_{\AIE}(\pi)} = \tau_{\AIE}(\pi)$}.
\end{theorem}

Now, although results on unbiased estimation are helpful, they do not provide a complete
picture of what good estimators for $\tau_{\ADE}(\pi)$ and $\tau_{\AIE}(\pi)$ should look like, and what kinds of
guarantees we should expect. The direct effect estimator \eqref{eq:HT-ADE} is further studied by \citet{savje2017average},
who provide bounds on its mean-squared error under a moderately sparse network interference model as in
\eqref{eq:network}; roughly speaking, they assume that the graph $E_{ij}$ has degree bounded on the order of $o(\sqrt{n})$.
\citet{li2020random} prove a central limit theorem for $\htau_{\ADE}(\pi)$ under network interference with a random
graph generative model. Meanwhile, in the case of very sparse graphs, i.e.,  when vertices in the graph have bounded degrees, the indirect
effect estimator $\htau_{\AIE}(\pi)$ could be studied using methods developed in \citet{aronow2017estimating} and
\citet{leung2020treatment}. However, in even moderately dense settings, \citet{li2020random} find that unbiased estimators
of indirect effects may have vary large variance and caution against their use; they also propose alternative estimators that
are more stable---again under a random graph generative model. To the best of our knowledge, efficiency theory for treatment
effect estimation under interference remains as of now uninvestigated.

\section{Proofs}
\label{section:proofs}

\subsection{Proof of Theorem \ref{theo:overall}}
We start with a slightly more formal form of the infinitesimal policy effect, 
\[\tau_{\NUDGE}(\pi) = \sum_{k=1}^n \frac{\partial}{\partial \pi_k'} \cb{\frac{1}{n}\sum_{i = 1}^n \EE[{\pi}']{ Y_i} }_{\pi' = \pi},\]
i.e., we take derivative with respect to $\pi'$ and evaluate at $\pi$. For index $i$, we can rewrite $\EE{Y_i}$ in terms of the potential outcomes: 
\begin{align*}
\EE[\pi']{Y_i} & = \sum_{w_{-k}}\sum_{w_{k}\in\{0,1\}}Y_i(w_k ;w_{-k})\PP[\pi'_{-k}]{W_{-k} = w_{-k}} \PP[\pi_k']{W_k = w_k}\\
& = \sum_{w_{-k}}\sum_{w_{k}\in\{0,1\}}Y_i(w_k ;w_{-k})\PP[\pi'_{-k}]{W_{-k} = w_{-k}} \cb{w_k\pi_k' + (1-w_k)(1-\pi_k')}. 
\end{align*}
The dependency of this term on $\pi_k'$ is clear in this form. Taking a derivative with respect to $\pi_k'$ and evaluating at $\pi_k$ yields
\begin{align*}
\frac{\partial}{\partial \pi_k'}\EE[\pi']{Y_i}\big|_{\pi' = \pi} & = \sum_{w_{-k}} \cb{Y_i(w_k=1; w_{-k}) - Y_i(w_k=0;w_{-k})} \PP[\pi_{-k}]{W_{-k} = w_{-k}}\\
& =  E_{\pi}\cb{Y_i(w_k=1;W_{-k})-Y_i(w_k=0;W_{-k})}. 
\end{align*}
If we sum over the index $i$ and $k$ and multiply the term by $1/n$, we get
\begin{align*}
\tau_{\NUDGE}(\pi)
& = \frac{1}{n}\sum_{k = 1}^n \sum_{i = 1}^n E_{\pi}\cb{Y_i(w_k=1;W_{-k})-Y_i(w_k=0;W_{-k})}
= \tau_{\AOE}(\pi).
\end{align*}

\subsection{Proof of Theorem \ref{theo:unbiased}}

For each pair of indices $i,j$, we can write $E_\pi\cb{ \frac{W_i Y_j}{\pi_i} - \frac{(1 - W_i) Y_j}{1 - \pi_i}}$ as
\begin{align*}
&\EE[\pi]{ \frac{W_i Y_j }{\pi_i}} - E_\pi\cb{ \frac{(1 - W_i) Y_j}{1 - \pi_i}}\\
& \quad \quad = E_\pi\cb{ \frac{W_i Y_j(w_i = 1; W_{-i})}{\pi_i}} - E_\pi\cb{ \frac{(1 - W_i) Y_j(w_i = 0; W_{-i}) }{1 - \pi_i}}\\
& \quad \quad = \EE[\pi]{\frac{W_i}{\pi_i}}E_{\pi}\cb{Y_j(w_i = 1; W_{-i})} - \EE[\pi]{\frac{1-W_i}{1-\pi_i}} E_{\pi}\cb{ Y_j(w_i = 0; W_{-i}) }\\
&\quad \quad = E_{\pi}\cb{Y_j(w_i=1;W_{-i})-Y_j(w_i=0;W_{-i}) }. 
\end{align*}
Thus summing over $i$ and $j$, we get
\begin{align*}
E_\pi\cb{\htau_{\AIE}(\pi)} &= \frac{1}{n} \sum_{i=1}^n \sum_{\{j \neq i:E_{ij}=1\}} E_\pi\cb{ \frac{W_i Y_j }{\pi_i} -  \frac{(1 - W_i) Y_j}{1 - \pi_i}}\\
&= \frac{1}{n}\sum_{i=1}^n \sum_{\{j \neq i:E_{ij}=1\}} E_{\pi}\cb{Y_j(w_i=1;W_{-i})-Y_j(w_i=0;W_{-i}) }. 
\end{align*}
When there is no edge connecting $i$ and $j$, the value of $Y_j$ does not depend on $w_i$, hence $Y_j(w_i=1;W_{-i}) = Y_j(w_i=0;W_{-i})$. Hence we can rewrite the above term:
\begin{align*}
E_\pi\cb{\htau_{\AIE}(\pi)} 
&= \frac{1}{n}\sum_{i=1}^n \sum_{j \neq i} E_{\pi}\cb{Y_j(w_i=1;W_{-i})-Y_j(w_i=0;W_{-i}) } 
= \tau_{\AIE}(\pi). 
\end{align*}

\subsection{Derivations for Example \ref{exam:1}}
It's clear from \eqref{exam1} that $Y_i(w_i = 1; W_{-i}) - Y_i(w_i = 0; W_{-i}) = \beta_2$ and $Y_j(w_i=1;W_{-i})-Y_j(w_i=0;W_{-i})  = \beta_3 E_{ij} /{\sum_{j \neq i} E_{ij}}$. Hence
\begin{align*}
&\tau_{\ADE} = \frac{1}{n} \sum_{i=1}^n E_{\pi}\cb{Y_i(w_i = 1; W_{-i}) - Y_i(w_i = 0; W_{-i})} 
= \beta_2,  \\
&\tau_{\AIE} = \frac{1}{n}\sum_{i = 1}^n \sum_{j \neq i}  E_{\pi}\cb{Y_j(w_i=1;W_{-i})-Y_j(w_i=0;W_{-i}) }
= \frac{1}{n}\sum_{i = 1}^n \sum_{j \neq i}  \p{\beta_3 E_{ij}/{\sum_{j \neq i} E_{ij}}} = \beta_3. 
\end{align*}

\subsection{Derivations for Example \ref{exam:2}}
Model \eqref{exam2} implies that $Y_i(w_i = 1; W_{-i}) - Y_i(w_i = 0; W_{-i}) = \beta_i$ and $Y_j(w_i=1;W_{-i})-Y_j(w_i=0;W_{-i})  = \nu_{ij}$. Hence 
\[
\tau_{\ADE} = \frac{1}{n} \sum_{i=1}^n \beta_i, \ \ \ \
\tau_{\AIE} = \frac{1}{n}\sum_{i = 1}^n \sum_{j \neq i}  E_{\pi}\cb{Y_j(w_i=1;W_{-i})-Y_j(w_i=0;W_{-i}) }
= \frac{1}{n}\sum_{i = 1}^n \sum_{j \neq i}  \nu_{ij}. 
\]

\subsection{Derivations for Example \ref{exam:3}}
In this clustered setup, the treatment assignment of units outside of the cluster has no effect on the unit's observed outcome.
With a slight abuse of notation, we use $W_{-i}$ to denote the treatment vector of all the units in the same cluster as unit $i$, except the unit itself.
To start with, we find the distribution of $W_{-i}$ marginalized over $W_i$ under this design, where
\[
P(W_{-i}=w)= 
\begin{cases}
    1-\rho+\frac{\rho}{m},& \text{if } w=(0,\dots,0)\\
    \frac{\rho}{m},& \text{if } w\in\{e_1,\dots,e_{m-1}\}\\
    0,              & \text{otherwise},
\end{cases}
\]
where $e_j\in\RR^{m-1}$ denotes the vector with a $1$ in the $j$th coordinate and $0$'s elsewhere. Thus, writing $C_i$ for the indicator that unit $i$ in a treated cluster and recalling that, in our randomization design, if $W_i = 1$ then the $i$-th unit cannot have a treated neighbor:
\begin{align*}
\tau_{\ADE}(\pi) &= \frac{1}{n} \sum_{i=1}^n E\cb{Y_i(w_i = 1; W_{-i}) - Y_i(w_i = 0; W_{-i})}\\
&= \frac{1}{n} \sum_{i=1}^n \left[ \PP{W_i = 0 \ \& \ C_i = 1}\cb{Y_i(\text{treated \& exposed}) - Y_i(\text{exposed})} \right.\\
&\qquad\qquad\left. + \cb{\PP{W_i = 1} + \PP{C_i = 0}} \cb{Y_i(\text{treated}) - Y_i(\text{none})} \right ] \\
&= \p{\rho-\frac{\rho}{m}}\tau_{\text{SELF},1}+\p{1-\rho+\frac{\rho}{m}}\tau_{\text{SELF},0}, \\ \\
\tau_{\AIE}(\pi) &= \frac{1}{n} \sum_{i=1}^n\sum_{j\ne i,E_{ji}=1} E\cb{Y_i(w_j = 1; W_{-j}) - Y_i(w_j = 0; W_{-j})}\\
&= \frac{1}{n} \sum_{i=1}^n \left[(m - 1)\PP{W_i = 1} \cb{Y_i(\text{treated \& exposed}) - Y_i(\text{treated})} \right.\\
&\qquad\qquad\left. + \cb{(m - 1)\PP{C_i = 1} + \PP{W_i = 0 \ \& \ C_i = 1}}\cb{Y_i(\text{exposed}) - Y_i(\text{none})} \right ] \\
&= (m-1)\cb{\frac{\rho}{m}\tau_{\text{SPILL},1}+\p{1-\rho+\frac{\rho}{m}}\tau_{\text{SPILL},0} }.
\end{align*}

\end{appendix} 


\end{document}